\documentclass[12pt,psfig,epsf]{article}
\usepackage{epsfig} \usepackage{amssymb} \usepackage{amsfonts}

\def\beq{\begin{equation}}
\def\eeq{\end{equation}}
\def\bea{\begin{eqnarray}}
\def\eea{\end{eqnarray}}
\def\bem{\begin{math}}
\def\eem{\end{math}}
\def\bit{\begin{itemize}}
\def\eit{\end{itemize}}
\def\bla{\begin{flushright}}
\def\ela{\end{flushright}}
\def\qq2{$Q^2$}               % Q2
\def\aa1{$A_1(x,Q^2)$}        % A1
\def\ff1{$F_1(x,Q^2)$}        % F1
\def\gg1{$g_1(x,Q^2)$}        % g1
\newcommand{\z}{&\hspace*{-8pt}}

\newpage
\newcommand{\df}[2]{\mbox{$\frac{#1}{#2}$}}

\newcommand{\la}{\lambda}
\newcommand{\De}{\Delta}
\newcommand{\rd}{{\rm d}}

\begin{document}
%---------------------------------------------------------------------

%                          Title
\begin{center}
{\Large \bf 
Some methods for the evaluation of complicated Feynman integrals
} \\

\vspace{4mm}

%                      author/address
A.V.Kotikov\\
Bogoliubov Theoretical Physics Laboratory,
%\\ 
Joint Institute for Nuclear Research \\
141980 Dubna, Russia.\\
\end{center}

%                        Abstract
\begin{abstract}

We discuss a progress in calculations of Feynman integrals 
%which has been done with help of 
based on the Gegenbauer Polynomial Technique and 
the Differential Equation Method. We
demonstrate the results for a class of two-point two-loop diagrams and 
the evaluation of most complicated part of ${\rm O}(1/N^3)$ contributions to 
critical exponents of $\phi^4$-theory.
An illustration of the results obtained with help of above methods is
considered.\\
%for any spacetime dimensionality $D$.\\

\end{abstract}
%\newpage

Last years there was an essential progress in calculations
of Feynman integrals. It seems that most important results have
been obtained for two-loop four-point massless Feynman diagrams:
in on-shall case (see \cite{Smirnov,GeRede}) and for a class of off-shall
legs (see \cite{Offshall}). A review of the results can be found in
\cite{GeReconf}. Moreover, very recently results for a class of these
diagrams have been obtained
\cite{mass} in the case when some propagators have a nonzero mass.\\

In the paper, I review two methods for calculations of Feynman diagrams. 

The
first one, so-called the Gegenbauer Polynomial Method (see \cite{2} and
also \cite{3}-\cite{4}), has been used in particular for the evaluation 
of $\alpha_s$-
corrections to the longitudinal structure function of deep inelastic 
scattering process. The structure of the results in Mellin moment space
(see \cite{3}-\cite{KazKo}) is very similar to the coefficients 
in \cite{Smirnov,mass} of the Mellin-Barnes transforms  for the above 
double-bokses.
The coefficients are similar also to ones which have arised
%have been appeared 
(see \cite{FKV1}-\cite{FKV2})
in expansions over the inversed
mass for some two-loop two-point and three-point diagrams. 

A version of the second method, which is called as the Differential Equation 
Method \cite{DEM1}-\cite{DEMrev}, 
has been used in above calculations (see \cite{GeRede,GeReconf}
and references therein).

An illustration of some results which have been obtained with help of 
these two methods is
considered.
The additional information about a modern progress in calculations
of Feynman integrals can be found, for example, also in recent articles 
\cite{Passarino,Laporta}.\\

\vspace{4mm}

{\Large \bf A.~~~The Gegenbauer Polynomial Technique}\\ 
%the evaluation of complicated Feynman integrals } \\

\section{Basic Formulae}

Fifteen years ago
the method based on the expansion of propagators in Gegenbauer series
(see \cite{1.1}) has been introduced in \cite{2,2.1}.
 One has shown  \cite{2,ChSm} that by this
method the analytical evaluation of counterterms in the minimal
subtraction scheme at the 4-loop level in any model and for any
composite operator was indeed possible. The Gegenbauer Polynomial (GP)
 technique has been applied
successfully for propagator-type Feynman diagrams (FD) in many calculations 
(see \cite{2,2.1}).
In the Section {\bf A}
%present paper 
we consider
{\it a 
development of the GP technique} (obtained in \cite{4}), an illustration
of the results obtained in \cite{BrKo} and the application
of the results calculated in \cite{3.1}.

Throughout the Section {\bf A}
%paper 
we use the following notation. The use of dimensional
regularization is assumed. All the calculations are performed in the
%coordinate 
space
 of dimension $D=4-2\varepsilon$.
Note that contrary to \cite{2} we
analyze FD directly in  momentum $x$-space which allows
us to avoid the appearance of Bessel functions.
%It is possible 
Because we consider here only
%in the case of 
propagator-type massless FD, we know
%because 
their
dependence on a single external momentum 
%(in $p$-space) or on a single external coordinate is power-like and known 
beforehand. The point of
interest is the coefficient function $C_f$, which depends on
$D=4-2\varepsilon$ and is a Laurent series in $\varepsilon $. \\

{\bf 1.1}~~
First of all,
%Here
we present useful formulae to use of Gegenbauer polynomials.
% and traceless products.
Following \cite{2,2.1}, $D$-space integration can
be represented in the form
$$
d^Dx~=~ \frac{1}{2} S_{D-1}(x^2)^{\lambda} dx^2 d\hat{x}~~~~
(\lambda = D/2-1),
$$
where $
%\bf{x}^2=\vec{x}^2,
{}~\hat{x}=\vec{x}/\sqrt{
%\vec{x}
x^2}$, 
%$\lambda = D/2-1=1-\varepsilon$
and
$S_{D-1}=2 {\pi}^{\lambda+1}/\Gamma(\lambda+1)$ is the surface of the
unit hypersphere in $R_D$. The Gegenbauer polynomials
$C_n^{\delta}(t)$ are defined as \cite{1.1,2.1}
\begin{eqnarray}
 (1-2rt + r^2)^{-\delta} ~=~ \sum^{\infty}_{n=0}~C_n^{\delta}(t)r^n
{}~~~~(r \leq 1),~~~~ C_{n}^{\delta}(1) ~=~ \frac{ \Gamma(n+2\delta)}
{n!\Gamma(2\delta)},
  \label{A1}
\end{eqnarray}
whence the expansion for the propagator is:
\begin{eqnarray}
\frac{1}{(x_1-x_2)^{2\delta}} ~=~
\sum^{\infty}_{n=0} ~C_n^{\delta}(\hat{x_1}\hat{x_2})~
\Biggl[
\frac{{(x^2_1)}^{n/2}}{{(x^2_2)}^{n/2 + \delta}} \Theta(x_2^2 - x_1^2)
{}~+~
\Bigl(x_1^2 \longleftrightarrow x_2^2  \Bigr)  \Biggr],
  \label{A2}
\end{eqnarray}
where
\[ \Theta(y)~=~ \left\{
\begin{array}{rl}
1, & \mbox{ if }y \geq 0 \\
0, & \mbox{ if }y  <   0
\end{array} \right.  \]

Orthogonality of the Gegenbauer polynomials $C_n^{\lambda}(x)$
%having the index $\lambda$
is expressed by the equation (see \cite{2})
\begin{eqnarray}
\int ~C_n^{\lambda}(\hat{x_1}\hat{x_2})~
 \, C_m^{\lambda}(\hat{x_2}\hat{x_3}) \, d\hat{x_2}
~=~\frac{\lambda}{n+\lambda}\, \delta^m_n
 ~C_n^{\lambda}(\hat{x_1}\hat{x_3}),~
  \label{A3}
\end{eqnarray}
where $\delta^m_n$ is the Kronecker symbol.

%To our study
The following formulae are useful (see \cite{2,3}):
\begin{eqnarray}
%\z
C_n^{\delta}(x)&=&
\sum_{p \geq 0} \frac{(2x)^{n-2p}(-1)^p \Gamma(n-p+\delta)}
{(n-2p)!p!\Gamma(\delta)}
%\label{A4}   \\\z
{}~~~
{}~ \mbox{ and }~
\nonumber   \\ 
%\z
\frac{(2x)^{n}}{n!} &=& \sum_{p \geq 0}
C_{n-2p}^{\delta}(x) \frac{(n-2p+\delta)\Gamma(\delta)}
{p!\Gamma(n-p+\delta+1)}  \label{A4}
\end{eqnarray}

Substituting the latter equation from (\ref{A4}) for $\delta = \lambda$ to
the first one, we have the following equation after the separate
analyses at odd and even $n$:
\begin{eqnarray}
C_n^{\delta}(x)~=~\sum_{k=0}^{[n/2]}
C_{n-2p}^{\lambda}(x) \frac{(n-2k+\lambda) \Gamma(\lambda)}
{k!\Gamma(\delta)}  \frac{ \Gamma(n+\delta-k)\Gamma(k+\delta-\lambda)}
{\Gamma(n+\lambda+1-k)\Gamma(\delta-\lambda)}  \label{A7}
\end{eqnarray}

\vskip 0.5cm

{\bf 1.2}~~
Following \cite{2,3} we introduce the traceless product (TP)
$x^{\mu_1...\mu_n}$ connected with the usual product
$x^{\mu_1}...x^{\mu_n}$ by the following equations 
%(see \cite{2,3})
\begin{eqnarray}
%\z 
x^{\mu_1...\mu_n}&=&\hat{S}
\sum_{p \geq 0} \frac{n!(-1)^{p} \Gamma(n-p+\lambda)}
{2^{2p} p! (n-2p)!\Gamma(n+\lambda)}~
g^{\mu_1\mu_2}...g^{\mu_{2p-1}\mu_{2p}}~x^{2p}~x^{\mu_{2p+1}}...x^{\mu_n}
%\label{B1}   \\
               \nonumber  \\
& & \nonumber  \\
%\z 
x^{\mu_1}...x^{\mu_n}&=&\hat{S}
\sum_{p \geq 0} \frac{n! \Gamma(n-2p+\lambda+1)}
{(2)^{2p} p! (n-2p)!\Gamma(n-p+\lambda+1)}~
g^{\mu_1\mu_2}...g^{\mu_{2p-1}\mu_{2p}}~x^{2p}~x^{\mu_{2p+1}...\mu_n}
\label{B1}
\end{eqnarray}
%where hereafter $\lambda \equiv D/2-1$.

Comparing Eqs.(\ref{A4})
%, (\ref{A6})
and (\ref{B1}), we obtain the
following relations between TP and GP
\begin{eqnarray}
z^{\mu_1...\mu_n}~x^{\mu_1...\mu_n}&=& \frac{n! \Gamma(\lambda)}
{2^n \Gamma(n+\lambda)}~C_n^{\lambda}(\hat{x}\hat{z})~{(x^2 z^2)}^{n/2},~
%~~\mbox{ and, hence, }~~
\nonumber \\
 x^{\mu_1...\mu_n}~x^{\mu_1...\mu_n}&=& 
\frac{\Gamma(n+2\lambda) \Gamma(\lambda)}
{2^n \Gamma(2\lambda)\Gamma(n+\lambda)}~x^{2n}
  \label{B2}
\end{eqnarray}
We give also the simple but quite useful conditions:
\begin{eqnarray}
z^{\mu_1...\mu_n}~x^{\mu_1...\mu_n}~=~
z^{\mu_1}...\,z^{\mu_n}~x^{\mu_1...\mu_n}~=~
z^{\mu_1...\mu_n}x^{\mu_1}...\,x^{\mu_n},
  \label{B3}
\end{eqnarray}
which
 follow immediately from the TP definition:
$g^{\mu_i\mu_j}~x^{\mu_1...\mu_i...\mu_j...\mu_n}~=~ 0$.\\

The use of the TP $x^{\mu_1...\mu_n}$ makes it possible to ignore
terms of the type $g^{\mu_i\mu_j}$ that arise upon integration: they can
be readily recovered from the general structure of the TP. Therefore,
in the process of integration it is only necessary to follow the
coefficient of the leading term $x^{\mu_1}...x^{\mu_n}$.
The rules to integrate FD containing TP can be found, for example in
\cite{3,3.1,3.2}. For a loop we have 
(hereafter $Dx \equiv (d^Dx)/(2\pi)^D$)
\footnote{The Eq.(\ref{4}) has been used in \cite{3,3.1,3.2,KazKo} for 
calculations of the moments of structure functions of deep inelastic 
scattering.}:
\begin{eqnarray}
\int Dx \frac{ x^{\mu_1...\mu_n}}{x^{2\alpha}(x-y)^{2\beta}}
{}~=~
\frac{1}{(4\pi)^{D/2}}~
\frac{y^{\mu_1...\mu_n}}{y^{2(\alpha + \beta - \lambda -1)}}
{}~ A^{n,0}(\alpha, \beta),
  \label{4}
\end{eqnarray}\\
where
$$A^{n,m}(\alpha,\beta)~=~
\frac{a_n(\alpha)a_m(\beta)}{a_{n+m}(\alpha+\beta-\lambda-1)}
~~\mbox{ and }~~
a_n(\alpha)~=~\frac{\Gamma (D/2-\alpha +n )}{\Gamma(\alpha)}
$$

Note that in our analysis it is necessary to consider more
complicate cases of integration, when the integrand  contains
$\Theta$ functions. Indeed, using the Eqs.(\ref{A2}) and (\ref{B2}),
we can represent
the propagator $(x_1-x_2)^{-2\lambda}$ into the following form
\footnote{In the case of the propagator $(x_1-x_2)^{-2\delta}$ with
$\delta \neq \lambda$ we should use also Eq.(\ref{A7}).}:
\begin{eqnarray}
\frac{1}{(x_1-x_2)^{2\lambda}} ~=~
\sum^{\infty}_{n=0} ~\frac{2^n \Gamma(n+\lambda)}{n!\Gamma(\lambda)}
{}~x_1^{\mu_1...\mu_n}~x_2^{\mu_1...\mu_n}~
\Biggl[
\frac{1}{x_2^{2(\lambda +n)}} \Theta(x_2^2 - x_1^2)
{}~+~
\Bigl(x_1^2 \longleftrightarrow x_2^2  \Bigr)  \Biggr]
  \label{B6}
\end{eqnarray}
%This equation was already applied in Section 3.

Using the GP properties from  previous subsection
and the connection (\ref{B2})
between GP and TP,
we obtain the  rules for calculating
FD with the $\Theta$-terms and TP.\\

{\bf 1.3}~~ {\it The rules}
have the following form:
\begin{eqnarray}
\int Dx \frac{x^{\mu_1...\mu_n}}{x^{2\alpha}(x-y)^{2\beta}}
~\Theta(x^2-y^2)
%\nonumber \\ \z
&=&
\frac{1}{(4\pi)^{D/2}}~
\frac{y^{\mu_1...\mu_n}}{y^{2(\alpha + \beta - \lambda -1)}}
~ \sum_{m=0}^{\infty}
\frac{B(m,n|\beta,\lambda)}{m+\alpha + \beta - \lambda -1}~ \nonumber \\
%\z 
&\stackrel{(\beta =
  \lambda)}{=}& \hspace{-0.3cm}
 \frac{1}{(4\pi)^{D/2}}~\frac{y^{\mu_1...\mu_n}}{y^{2(\alpha  -1)}}
{}~\frac{1}{\Gamma(\lambda)}
\frac{1}{(\alpha -1)(n+\lambda)}  \label{2} \\
\int Dx \frac{ x^{\mu_1...\mu_n}}{x^{2\alpha}(x-y)^{2\beta}}
~\Theta(y^2-x^2)
%\nonumber \\ \z
 &=&
\frac{1}{(4\pi)^{D/2}}~
\frac{y^{\mu_1...\mu_n}}{y^{2(\alpha + \beta - \lambda -1)}}
{}~ \sum_{m=0}^{\infty}
\frac{B(m,n|\beta,\lambda)}{m+n- \alpha + \lambda +1}~  \nonumber \\
%\z 
&\stackrel{(\beta =
  \lambda)}{=}& \hspace{-0.3cm}
 \frac{1}{(4\pi)^{D/2}}~\frac{y^{\mu_1...\mu_n}}{y^{2(\alpha -1)}}
{}~\frac{1}{\Gamma(\lambda)}
\frac{1}{(n+\lambda +1- \alpha )(n+\lambda)}  \label{3}
\end{eqnarray}\\
where
$$B(m,n|\beta,\lambda)~=~\frac{\Gamma(m+\beta +n)}
{m! \Gamma(m+n+1+\lambda) \Gamma(\beta)}~
\frac{\Gamma(m+\beta - \lambda)}{\Gamma(\beta - \lambda)}~~
%\mbox{ and }~~ \lambda=D/2-1
$$

The sum of above diagrams does not contain $\Theta$-terms
and should reproduce Eq.(\ref{4}).
To compare the r.h.s. of Eqs.(\ref{2},\ref{3}) and the r.h.s. of
Eq.(\ref{4}) we use the
transformation of $_3F_2$-hypergeometric function with unit argument
${}_3F_2(a,b,c;e,b+1;1)$
(see \cite{5}):

\begin{eqnarray}
%\z
 \sum_{k=0}^{\infty} \frac{\Gamma(k+a)\Gamma(k+c)}
{k!\Gamma(k+f)} \frac{1}{k+b}
=\frac{\Gamma (a)\Gamma(1-a)\Gamma(b)\Gamma(c-b)}
{\Gamma(f-b)\Gamma(1+b-a)} \nonumber  \\
- \frac{\Gamma(1-a)\Gamma(a)}{\Gamma(f-c)\Gamma(1+c-f)}\,
%\cdot~
 \sum_{k=0}^{\infty} \frac{\Gamma(k+c-f+1)\Gamma(k+c)}
{k!\Gamma(k+1+c-a)} \frac{1}{k+c-b}
  \label{8.2}
\end{eqnarray}
This is the case (when $k=m, b=\alpha+\beta-\lambda-1 , c=n+\beta$) to
compare Eq.(\ref{4}) and the sum of Eqs.(\ref{2},\ref{3}).

Analogously to Eqs.(\ref{2}) and (\ref{3}) we have more
complicate cases:
%\footnote{The full set of rules will be presented in  \cite{4}.}:

\begin{eqnarray}
& &\int Dx \frac { x^{\mu_1...\mu_n}}{x^{2\alpha}(x-y)^{2\beta}}
\Theta(x^2-z^2)
      ~=~ \frac{1}{(4\pi)^{D/2}}~
y^{\mu_1...\mu_n}
\Biggr[
\frac{\Theta (y^2 -z^2)}{y^{2(\alpha + \beta - \lambda -1)}}
{}~ A^{n,0}(\alpha,\beta)
 \nonumber \\ 
& & + ~
 \sum_{m=0}^{\infty}
~\frac{B(m,n|\beta,\lambda)}{z^{2(\alpha + \beta - \lambda -1)}}
 ~ \biggl( {\Bigl(\frac{y^2}{z^2}\Bigr)}^{m}
\frac{\Theta (z^2 -y^2)}{m+\alpha + \beta - \lambda -1}~-~
{\Bigl(\frac{z^2}{y^2}\Bigr)}^{m+ \beta +n}
\frac{\Theta (y^2 -z^2)}{m-\alpha +n+1+ \lambda } \biggr)
 \Biggr]
                         ~ \nonumber \\
& & \hspace{2cm}
\stackrel{(\beta =
  \lambda)}{=}
 \frac{1}{(4\pi)^{D/2}}~ \frac{1}{\Gamma(\lambda)}~y^{\mu_1...\mu_n}~
\Biggl[
\frac{1}{y^{2(\alpha -1)}}
\frac{\Theta (y^2-z^2)}{(\alpha -1)(n+\lambda +1 -\alpha)}  \nonumber \\
& &\hspace{2.5cm} 
+~ \frac{1}{z^{2(\alpha -1)}} \frac{1}{n+\lambda }
\biggl(
\frac{\Theta (z^2-y^2)}{\alpha -1}~-~
{\Bigl(\frac{z^2}{y^2}\Bigr)}^{n+ \lambda }
\frac{\Theta (y^2-z^2)}{n+1+ \lambda -\alpha } \biggr)
%]
 \Biggr]  \label{5}\\
& &  \nonumber  \\
& & \int Dx \frac { x^{\mu_1...\mu_n}}{x^{2\alpha}(x-y)^{2\beta}}
\Theta(z^2-y^2)
{}=~ \frac{1}{(4\pi)^{D/2}}~
y^{\mu_1...\mu_n}~
\Biggl[
\frac{\Theta (z^2 -y^2)}{y^{2(\alpha + \beta - \lambda -1)}}
{}~A^{n,0}(\alpha,\beta)
 \nonumber \\ 
& & - ~  \sum_{m=0}^{\infty}
{}~\frac{B(m,n|\beta,\lambda)}{z^{2(\alpha + \beta - \lambda -1)}}
{}~\biggl( {\Bigl(\frac{y^2}{z^2}\Bigr)}^{m}
 \frac{\Theta (z^2 -y^2)}{m+\alpha + \beta - \lambda -1}~-~
{\Bigl(\frac{z^2}{y^2}\Bigr)}^{m+ \beta +n}
\frac{\Theta (y^2 -z^2)}{m-\alpha +n+1+ \lambda } \biggr)
 \Biggr]
{}~\nonumber \\
& & \hspace{2cm} 
\stackrel{(\beta =
  \lambda)}{=}
 \frac{1}{(4\pi)^{D/2}}~ \frac{1}{\Gamma(\lambda)}~y^{\mu_1...\mu_n}~
\Biggl[
\frac{1}{y^{2(\alpha -1)}}
\frac{\Theta (z^2-y^2)}{(\alpha -1)(n+\lambda +1 -\alpha)}  \nonumber \\
& & \hspace{2.5cm}
 -~ \frac{1}{z^{2(\alpha -1)}} \frac{1}{n+\lambda }
\biggl(
\frac{\Theta (z^2-y^2)}{\alpha -1}~-~
{\Bigl(\frac{z^2}{y^2}\Bigr)}^{n+ \lambda }
\frac{\Theta (y^2-z^2)}{n+1+ \lambda -\alpha } \biggr)
 \Biggr]  \label{6}
\end{eqnarray}\\
One can easily see that the sum of the above diagrams lead to
results identical to (\ref{4}).\\

%\section{The evaluation of a class of Feynman diagrams}

\section{Calculation of complicated FD}

% {\bf 2.} 
{\it The
aim of this section} is to demonstrate the result of \cite{4} for
%study  
a class of master
two-loop diagrams containing the vertex with two propagators having index
1 or $\lambda$.

%{\bf 1.}
Consider 
%in the $x$-space 
the following general diagram

$$ \int  \frac {Dx Dy}{y^{2\alpha}(z-y)^{2t}(z-x)^{2\beta}x^{2\gamma}
(x-y)^{2s}}
{}~\equiv ~J(\alpha,t, \beta, \gamma, s)$$
and restrict ourselves to the FD
$A(\alpha, \beta, \gamma ) ~=~ J(\alpha,\lambda, \beta, \gamma,
\lambda)$,
which is the one
of FD of interest for us here. It is easily
shown
(see \cite{VPH,3.2,4})
that ($\sigma = 3+ \lambda -(\alpha + \beta + \gamma )$)

\begin{eqnarray}
C_f[A(\alpha,\beta, \gamma)]~=~C_f[A(\alpha,\sigma, \gamma)]~=~
C_f[J(\gamma,\lambda, \lambda,\sigma, \alpha)]~=~
C_f[J(\sigma,\gamma,\lambda, \lambda,\beta)],   \label{7}
\end{eqnarray}
%{\bf 2.}
 Doing Fourier transformation of both: the diagram $A(\alpha,\beta,\gamma)$
%from eq.(\ref{8})
and its solution
in the form $C_f[A(\alpha, \beta, \gamma)](z^2)^{-\tilde \sigma}$,
where hereafter
$\tilde t = \lambda +1-t,~t= \{\alpha,\beta, \gamma, \sigma , ...\}$, and
considering the new
diagram as one in the momentum
$x$-space 
%(i.e. making the dual transformation)
we obtain the relation

\begin{eqnarray} 
%\z
C_f[A(\alpha,\beta, \gamma)] &=&
%&\stackrel{f}{=}&
\frac{a_0^2(\lambda)a_0(\alpha)a_0(\beta)a_0(\gamma)}{a_0(\delta)}~
C_f[J(\tilde \alpha,1,\tilde \beta,\tilde \gamma ,1)] 
\nonumber\\ 
%\z
%& \stackrel{d}{=}&
&=&
a_0^2(\lambda)a_0(\alpha)a_0(\beta)a_0(\gamma)a_0(\sigma)~
C_f[J(\tilde \beta,1,\tilde \alpha,\tilde \gamma,1)]~  \label{9}
\end{eqnarray}
between the diagram, which contains the vertex
with two propagators
having the index $\lambda$, and the similar diagram containing the vertex
with two 
propagators having the index 1.

%3.
Repeating the manipulations of 
%e would (see 
\cite{VPH,Kaz1,3.2,4}
%subsection 1 
we can obtain the
following relations:

\begin{eqnarray}
%\z
C_f[J(\tilde \beta,1,\tilde \alpha,\tilde \gamma,1)]=
C_f[J(\tilde \beta,1,\tilde \sigma,\tilde \gamma,1)]
%\nonumber \\ \z
=C_f[J(1,1,\tilde \gamma ,\tilde \sigma,\tilde \alpha )]=
C_f[J(\tilde \sigma,1,1,\tilde \gamma,\tilde \beta)]  \label{10}
\end{eqnarray}
Thus, we have obtained the relations between all diagrams from the class
introduced in the beginning of this section. Hence, it is necessary
to find the solution for one of them. We prefer to analyze the
diagram $A(\alpha,\beta,\gamma)$, that is the content of the next 
subsection.\\

%\section{The calculations}
{\bf 2.1}~~
{\it We calculate the diagram} $A(\alpha, \beta, \gamma)$ by the
following way\footnote{The symbol $\stackrel{(n)}{=}$ marks the fact
  that the equation $(n)$ is used on this step.}:

\begin{eqnarray}
%\z 
& & A(\alpha, \beta, \gamma) \stackrel{(\ref{B6})}{=}
% \mbox{ from Eq.(\ref{B6}) }=~
\sum_{n=0}^{\infty} \frac{2^n \Gamma(n+\lambda)}{n! \Gamma(\lambda)}
\int Dx Dy \frac { z^{\mu_1...\mu_n}}{x^{2\gamma}(z-x)^{2\beta}}
 \frac { y^{\mu_1...\mu_n}}{y^{2\alpha}(x-y)^{2\lambda}}
\cdot \nonumber \\
& &
\Bigl[\frac{\Theta(z^2-y^2)}{z^{2(n+\lambda)}}~+~
\frac{\Theta(y^2-z^2)}{y^{2(n+\lambda)}} \Bigr]~
%\mbox{ from eq.(\ref{5},\ref{6}) }=~
\label{11} \\
% \nonumber \\
& & 
%\hspace{0.5cm} 
 \stackrel{(\ref{5},\ref{6})}{=}
 \frac{1}{(4\pi)^{D/2}}~ \frac{1}{\Gamma(\lambda)}~\frac{1}{\alpha -1}
\sum_{n=0}^{\infty} \frac{2^n \Gamma(n+\lambda)}{n! \Gamma(\lambda)}
\int Dx \frac {z^{\mu_1...\mu_n}x^{\mu_1...\mu_n} }{x^{2\gamma}(z-x)^{2\beta}}
\cdot \nonumber \\
& & \Biggl[
\frac{1}{\lambda +n+1-\alpha}  
\bigl(
\frac{\Theta(z^2-x^2)}{z^{2(n+\lambda)}x^{2(\alpha -1)}}~+~
\frac{\Theta(x^2-z^2)}{x^{2(n+\lambda)}z^{2(\alpha -1)}} \bigr)
\nonumber \\  
%\z
& & \hspace{4cm} ~-~
\frac{1}{\lambda +n+\alpha -1} \cdot \bigl(
\frac{\Theta(z^2-x^2)}{z^{2(n+\lambda +\alpha -1)}}~+~
\frac{\Theta(x^2-z^2)}{x^{2(n+\lambda +\alpha -1) }} \bigr)
\Biggr]  \nonumber 
%\label{11}
\end{eqnarray}

After some algebra we have got (see \cite{4}) the result in the form:
$$ C_f[A(\alpha, \beta, \gamma)]~=~ \frac{1}{(4\pi)^{D}}~
\frac{1}{\Gamma(\lambda)}~
\frac{1}{\alpha -1}~
\Bigl[
\overline I ~-~ \tilde I
\Bigr],$$
where
\footnote{We would like to note that the coefficients in 
Eqs.(\ref{13}) and (\ref{18})
are similar to ones (see \cite{FKV})
%which 
appeared in calculations of FD with massive propagators having the mass $m$. 
%The coefficients in front of $(z^2/m^2)^n$
The representation of the results for these diagrams in the form 
$\sum \varphi_n (z^2/m^2)^n$ ($ \varphi_n $ are the coefficients, which
are similar to ones in Eqs.(\ref{13}) and (\ref{18}))
is very convenient to obtain the results for more complicated FD by 
integration in respect of $m$ (see \cite{DEM1}-\cite{DEMrev}) of results less 
complicated FD.}

%$$
\begin{eqnarray}
%\z 
\overline{I}&=&
\sum_{n=0}^{\infty}
\frac{\Gamma(n+2 \lambda )
%a_n(2\lambda)
}{n! \Gamma (2 \lambda) }~
\Biggl[
\frac{1}{\lambda +n+1-\alpha} \cdot \biggl(
A^{n,0}(\alpha -1+\gamma, \beta )+A^{n,0}(n+\lambda +\gamma, \beta )
\biggr)
      \nonumber \\ 
%\z 
&-& \frac{1}{\lambda +n+\alpha -1} \cdot \biggl(
A^{n,0}(\gamma ,\beta )+A^{n,0}(n+\alpha + \lambda +\gamma -1, \beta)
\biggr) \Biggr]  \label{13}\\
& &     \nonumber \\ 
\tilde{I} &=&
 \frac{ \Gamma(1-\beta)\Gamma(\lambda +1-\alpha)
\Gamma(\lambda -1+  \alpha)\Gamma(1-\beta +\lambda)\Gamma(1- \gamma)
\Gamma(\alpha+\beta+\gamma -\lambda -2)}{\Gamma(2\lambda)
\Gamma(2+\lambda-\alpha -\beta)\Gamma(\alpha +\gamma -1)
\Gamma(2+\lambda-\gamma-\beta)\Gamma(\alpha +\beta -\lambda-1)}
       \nonumber  \\
%\z 
&-&    \sum_{n=0}^{\infty}
\frac{\Gamma(n+2 \lambda )}{n! \Gamma (2 \lambda) }~
\frac{(-1)^n \Gamma(1-\beta)}{\Gamma(\beta -\lambda)} \cdot
\frac{1}{\lambda +n+\alpha -1}   \label{18}  \\
%[
%\z 
&\times&   \Biggl[
 \frac{\Gamma(\alpha+\beta+\gamma -\lambda -2)
\Gamma(2-\alpha  -\gamma )}{\Gamma(3-\alpha -\beta -\gamma -n)
\Gamma(\alpha +\gamma +\lambda-1+n)} \nonumber\\
& & \hspace{5cm}~+~
%                            \nonumber \\ \z
\frac{\Gamma(1-\gamma)\Gamma(\beta +\gamma  -\lambda -1)}
{\Gamma(\gamma-\lambda -n)
\Gamma(2+2\lambda-\beta-\gamma+n)}
%]
\Biggr]  \nonumber
\end{eqnarray}

Thus, a quite simple solution for $A(\alpha, \beta, \gamma)$ is
obtained\footnote{ Before our studies, the possibility to represent
  $C_f[A(\alpha, \beta, \gamma)]$ as a combination of
  $_3F_2$-hypergeometric functions with unit argument, has been
  observed in \cite{Bro}.}. In next section we will consider the
important special case of
 these results.\\

{\bf 2.2}~~ {\it As a simple but important example} to apply these
results we consider the diagram $J(1,1,1,1,\alpha )$. It arises in the
framework of a number of calculations (see 
\cite{VPH,Kaz,Vas,Gra,Fadin}).
%Gri}, cite{Va}, \cite{Vas} and- \cite{Gra}). 
Its coefficient
function $I(\alpha )
\equiv C_f[J(1,1,1,1,\alpha )]$ can be found (see \cite{4}) as follows

\begin{eqnarray} 
%\z
I(\alpha ) &=&
\frac{a_0^4(1)a_0(\alpha )}{a_0(\alpha +2-2\lambda)}~
C_f[J( \lambda,\lambda,\lambda,\lambda,\tilde \alpha )]~~~~~ \mbox{ and }~~
\nonumber \\
& & C_f[J( \lambda,\lambda,\lambda,\lambda,\tilde \alpha )] =
C_f[A(\tilde \alpha ,3-\lambda -\tilde \alpha , \lambda)]
% \nonumber\\ \z
\nonumber
\end{eqnarray}
%The latter equation may be obtained by analogy with (\ref{10}).

{}From Eqs. (\ref{13}) and (\ref{18}) we obtain

   \begin{eqnarray}
%\z 
& & I(\alpha )~=~ - \frac{2}{(4\pi)^D}
 \frac{ \Gamma ^2(\lambda)\Gamma(\lambda - \alpha )
\Gamma(\alpha +1 -2 \lambda )}{\Gamma(2\lambda)
\Gamma(3\lambda -\alpha  -1)}
       \label{4.1}  \\
%\z 
& & \hspace{0.5cm} 
\times \Biggl[
\frac{ \Gamma ^2(1/2)\Gamma(3\lambda - \alpha -1)\Gamma(2\lambda - \alpha )
\Gamma(\alpha +1 -2\lambda )}{\Gamma(\lambda)
\Gamma(2\lambda +1/2 -\alpha )\Gamma(1/2-2\lambda +\alpha )}
\nonumber \\
& & \hspace{5cm}
~+~   \sum_{n=0}^{\infty}
\frac{\Gamma(n+2 \lambda )}{ \Gamma (n+\alpha +1) }~
\frac{1}{n+1 -\lambda +\alpha }
\Biggl]   \nonumber
\end{eqnarray}
Note that in \cite{Kaz} Kazakov has got another result for $I(\alpha )$:

   \begin{eqnarray}
%\z 
& &I(\alpha )~=~ - \frac{2}{(4\pi)^D}
 \frac{ \Gamma ^2(\lambda)\Gamma(1-\lambda)\Gamma(\lambda - \alpha )
\Gamma(\alpha +1 -2 \lambda )}{\Gamma(2\lambda)\Gamma(\alpha )
\Gamma(3\lambda -\alpha  -1)}
       \label{4.2}  \\
%\z 
& & 
%\hspace{0.5cm} 
\times \Biggl[
\frac{ \Gamma(\lambda)\Gamma(2- \lambda)\Gamma(\alpha )
\Gamma(3\lambda -\alpha  -1)}{
\Gamma(2\lambda -1)\Gamma(3-2\lambda )} \nonumber \\
& &\hspace{2cm}
~-~   \sum_{n=0}^{\infty} (-)^{n}
\frac{\Gamma(n+2 \lambda )}{ \Gamma (n+2-\lambda) }~ \biggl(
\frac{1}{n+1 -\lambda +\alpha } ~+~ \frac{1}{n+2\lambda -\alpha } \biggr)
\Biggl]   \nonumber
\end{eqnarray}

{}From Eqs. (\ref{4.1}) and (\ref{4.2}) we obtain the
transformation rule for $_3F_2$-hypergeometric function with argument $-1$:
\begin{eqnarray}
%\z
& & {}_3F_2(2a,b,1;b+1,2-a;-1)~=~b \cdot
\frac{\Gamma(2-a)\Gamma(b+a-1)\Gamma(b-a)
\Gamma(1+a-b)}
{\Gamma(2a)\Gamma(1+b-2a)}
\label{4.3}  \\
%\z 
& & \hspace{2cm}
- \frac{1-a}{b+a-1} \cdot {}_3F_2(2a,b,1;b+1,b+a;1) \nonumber \\
& & \hspace{2cm}
- \frac{b}{1+a-b} \cdot {}_3F_2(2a,1+a-b,1;2+a-b,2-a;-1),
  \nonumber
\end{eqnarray}
where $a=\lambda$ and $b=1-\lambda + \alpha $ are used.

Equation (\ref{4.3}) has been explicitly checked at $a=1$ and $b=2-a$
(i.e. $\lambda =1$ and $\alpha =1$), where the
$_3F_2$-hypergeometric functions may be calculated
exactly. 
%We cannot directly 
It is very difficult to prove Eq.(\ref{4.3}) at arbitrary $a$ and $b$
values: the general proof seems to be non-trivial.
Note that it is different from the equations of \cite{5,PBM} and
may be considered
as a new transformation rule.

\section{Applications}

{\bf 3.1}~~~The above results have been used for evaluation
%calculations 
of very complicated FD
which contribute mostly in calculations based on
%in 
various type of $1/N$ expansions: 
%to the following results:
\begin{itemize}
\item
In the calculation (in \cite{QED3}) of the next-to-leading (NLO)
corrections to the value of dynamical mass generation (see \cite{Nash})
in the framework of three-dimensional Quantum Electrodynamics. 
\item
In the evaluation (in \cite{KoKo}) of the correct value of 
of the leading order contribution to the $\beta$-function
of the $\theta$-term in Chern-Simons theory. The $\beta$-function
is zero in the framework of usual perturbation theory but it takes nonzero
values in $1/N$ expansion (see \cite{korea}).
\item
In the evaluation (in \cite{Fadin}) of NLO corrections to the value of
gluon Regge trajectory (see discussions in \cite{Fadin} and references 
therein).
\item
In the calculation (in \cite{KoLi}) of the next-to-leading 
corrections to the BFKL intercept of spin-dependent part of high-energy
asymptotics of hadron-hadron cross-sections. 
\item
In the calculation (in \cite{KoLi,KoLi1}) of the next-to-leading 
corrections to the BFKL equation at arbitrary conformal spin.
\item
In the evaluation (in \cite{BrKo}) of the 
most complicated parts of ${\rm O}(1/N^3)$ contributions to 
critical exponents of $\phi^4$-theory,
for any spacetime dimensionality $D$.
\end{itemize}

We consider here only basic  steps of the last analysis \cite{BrKo}.
Since the pioneering work of the St Petersburg group~\cite{VPH,Imu},
exploiting conformal invariance~\cite{AMP} of critical phenomena,
it was known that the ${\rm O}(1/N^3)$ terms 
%$\eta_3$
in the large-$N$ critical exponents 
%$\eta$ 
of the non-linear $\sigma$-model,
or equivalently $\phi^4$-theory, in any number $D$ of
spacetime dimensions, derives its maximal complexity
from a single Feynman integral $I(\lambda)$ (see \cite{Imu}):
%The definition of $I$ is~\cite{Imu}
\begin{eqnarray}
I(\lambda)=\left.\frac{\rd}{\rd\De}\ln\Pi(\la,\De)\right|_{\De=0},
\label{def}
\end{eqnarray}
where
\begin{eqnarray}
\Pi(\la,\De)=\frac{{x^{2(\la+\De)}}}{\pi^D}\int\int\frac{\rd^D y \rd^D z}
{y^2z^2(x-y)^{2\la}(x-z)^{2\la}(y-z)^{2(\la+\De)}}\label{Pi}
\end{eqnarray}
is a two-loop two-point integral, with three dressed propagators,
made dimensionless by the appropriate power of $x^2$.

The result, obtained by GP technique, is
\begin{eqnarray}
I(\la)=\Psi(1)-\Psi(1-\la)+\frac{\Phi(\la)-\df13\Psi^{\prime\prime}(\la)
-\df{7}{24}\Psi^{\prime\prime}(1)}{\Psi^\prime(1)-\Psi^\prime(\la)}\,,
\label{ans}
\end{eqnarray}
where $\Psi(x)=\Gamma^\prime(x)/\Gamma(x)$ 
\footnote{The function $\Phi(\la)$ is quite similar to most complicated
part of the NLO corrections to BFKL intercept (see 
\cite{FadLip,KoLi,KoLi1} and references therein).} 
and 
\begin{eqnarray}
\Phi(\la)=4\int^1_0 d x \frac{x^{2\la-1}}{1-x^2}
\left\{{\rm Li}_2(-x)-{\rm Li}_2(-1)\right\}\,
~~~({\rm Li}_2(x)=\sum_{n>0}x^n/ n^2)
,\label{Phi}
\end{eqnarray}

In \cite{BGK,BrKo} the integral $I(\la)$ has been expanded near
$D=2$ and $D=3$, respectively, (i.e. for $D=2-2\varepsilon$ and 
$D=3-2\varepsilon$)
up to $\varepsilon^8$ in the form of alternative and non-alternative
double Euler sums \cite{LE,DZ1}.\\

%\newpage
{\bf 3.2}~~~The Eq. (\ref{4}) together with the ``uniqueness'' relations
\cite{VPH,Kaz1,Kaz} and the integration by parts \cite{IntPart,VPH} 
extended in \cite{3}-\cite{3.2} for the form of massless propagators 
%the form
$\sim x^{\mu_1,...,\mu_n}/(x^2)^{a}$ has been used for the evaluation
of the $\alpha_s$-corrections to the longitudinal structure function $F_L$
of deep-inelastic scattering process
\footnote{In the framework of supersymmertic extension the evaluation
of gluino contribution to $F_L$ has been done in \cite{gluino}.}. 
%Note that 
The corresponding results 
\cite{KaKoYaF,3,KazKo}
\footnote{We would like to note that the results \cite{3.1,PRL} contain
an error in gluon sector, which is not essential for phenomenology.
The correct results have been found in \cite{Neerven,KazKo}.}
contain the sums
\begin{eqnarray}
K_2(n)&=&\Bigl(1-(-1)^n \Bigr)\frac{1}{2}\zeta(2) +(-1)^n
\sum^{n}_{m=1}\frac{(-1)^{m+1}}{m^2},\nonumber \\
K_3(n)&=&\Bigl(1-(-1)^n \Bigr)\frac{3}{4}\zeta(3) +(-1)^n
\sum^{n}_{m=1}\frac{(-1)^{m+1}}{m^3},~~\nonumber \\
K_{2,1}(n)&=&\Bigl(1-(-1)^n \Bigr)\frac{5}{8}\zeta(3) +(-1)^n
\sum^{n}_{m=1}\frac{(-1)^{m+1}}{m^2}S_1(m)
\label{1.1}
\end{eqnarray}
which can be obtained by direct calculations (with help of the optical 
theorem
\footnote{The optical theorem is very powerful in calculations with
massive particles, too (see \cite{KoLiPaZo} and references therein).})
only at even $n$. The results for the anomalous dimensions of Wilson
operators contain also the function (\ref{1.1}) (see \cite{FlKoLa}).

The analytical continuation of
the results (\ref{1.1}) to integer values and to real ones (and even
to complex ones)
can be found, respectively, in \cite{3.1} and \cite{Ko94}. 
The continuation to the integer values
has been wide used in fits of deep inelastic experimental data at 
the next-to-leading-order (NLO)
approximation \cite{Kri,Vovk,KKDIS} and at 
the next-next-to-leading-order (NNLO) level 
\cite{PKK}-\cite{KPS}. The continuation to the real values
is very important for small Bjorken $x$ phenomenology. In the Ref.
\cite{Ko94} the extension of previous results \cite{LoYn,VoKoMa,
JeKoPa} has been performed for an approximation of Mellin
convolution by a sum of usual products. The extension give a possibility 
to obtain the following results:
\begin{itemize}
\item
To extend (in \cite{Q2evo}) the solution of DGLAP equation in 
double-logarithmic
approximation (see \cite{Rujula}) to NLO approximation.
\item
To explanate (in \cite{Ko96}) a sharp change of the ``intercept'' value 
$\alpha_{P}$ at $Q^2 \sim$ 2 GeV$^2$ for the power-like asymptotics 
of parton distributions
and structure function $F_2$, observed in \cite{AbraLevin}.
\item
To demonstrate (in \cite{FLR}) the positivity of the value for
longitudinal structure 
function of deep-inelastic scattering at small $x$ range. The results
have been obtained in so-called renormalization-invariant perturbation
theory (see Ref. \cite{Vovk} and references therein), which resums 
properly the large and negative NLO corrections (see \cite{Keller})
in the gluon part of 
the longitudinal Wilson coefficient function.
\item
To extract at small $x$ values the gluon density in \cite{KoPaGL}
(following to the articles \cite{Prytz}) and the longitudinal structure 
function $F_L$ (in \cite{KoPaFL}) from the experimental data for the 
structure function $F_2$ and its derivation $dF_2/d\ln{Q^2}$.
\end{itemize}

\vskip 0.9cm

{\Large \bf B.~~~The Differential Equation Method.}\\

\vskip 0.5cm
%\vskip 1.5cm
The idea of the Differential Equation Method 
(DEM) 
(see \cite{DEM1}-\cite{DEM3} and
%) (see a 
reviews in \cite{DEMrev}): 
to apply the integration by parts procedure \cite{IntPart,VPH} to an 
internal $n$-point subgraph
of a complicated Feynman diagram and later to represent new complicated
diagrams, obtained here, as derivatives in respect of corresponding masses 
of the initial diagram.

The integration by parts procedure \cite{IntPart,VPH}
%These recurrence relations are some particial cases of the relation
(see also \cite{DEM1}-\cite{DEMrev})
 for a general $n$-point (sub)graph with masses of its lines
$m_1, m_2, ..., m_n$, line momenta $p_1, p_2=p_1 - p_{12}, p_n=p_1 - p_{1n}$
and indices $j_1, j_2, .., j_n$, respectively, has the following form:
\begin{eqnarray}
%\z 
0&=& \int d^Dp_1 \frac{\partial}{\partial p_1^{\mu}}~ 
\Biggl\{p_1^{\mu} ~
{\biggl(\prod_{i=1}^{n}
c_i^{j_i} \biggr)}^{-1} \Biggr\} \label{1} \\ 
&=& \int d^Dp_1  {\biggl(\prod_{i=1}^{n}
c_i^{j_i} \biggr)}^{-1} \Biggl[ D - 2j_1 \biggl( 1- \frac{m_1^2}{c_1} \biggr) 
- \sum_{i=2}^{n} j_i 
 \biggl( 1-
\frac{m_1^2 + m_i^2 + p^2_{1i}-c_1}{c_i} \biggr)
\Biggr],
\nonumber
\end{eqnarray}
where
$c_k=p^2_k+m^2_k$ are the propagators of $n$-point (sub)graph. 

Because the diagram with the index $(j_i+1)$ of the propagator $c_i$ may be 
represented as the derivative (on the mass $m_i$), Eq.(\ref{1}) leads to 
the differential equations (in principle, to partial differential equations)
for the initial diagram (having the index $j_i$, respectively). This approach
which is based on
 the Eq.(\ref{1}) and allows to construct the (differential) relations
between diagrams has been named as Differential Equations Method (DEM).  
 For most
interested cases (where the number of the masses is limited)
these partial differential equations may be represented through
original differential equation\footnote{The example of the direct application
of the partial differential equation may be found in \cite{FJ1}.},
which is usually simpler to analyze.

Thus, we have got the differential equations for the initial diagram. The
inhomogeneous term contains only more simpler diagrams.
These simpler diagrams have more trivial topological structure 
and/or less number of loops \cite{DEM1} and/or ends \cite{DEM2,DEM3}.

Applying the procedure several times, we will
able to represent complicated Feynman integrals and their derivatives
(in respect of internal masses) through a set of quite simple well-known 
diagrams. 
Then, the results for the
complicated FD can be obtained by integration several times
of the known results for corresponding simple diagrams
\footnote{In calculations of real processes 
(essentially in the framework of Standard Model)
it is useful to use
the relation (1) (at least, at first steps of calculations)
to decrease the number of contributed diagrams (see
\cite{DEM1}-\cite{DEM3} and \cite{FTT} and references therein).}.

Sometimes it is useful (see \cite{Remiddi}) to use external momenta 
(or some their
functions) but not masses as parameters of integration.\\

%\newpage
\vskip 0.5cm

{\bf The recent progress in calculation of Feynman integrals with help 
of the DEM.} \\

{\bf 1.} The articles \cite{FKV1} and \cite{FKV}: \\

{\bf a)} The set of two-point two-loop FD with one- and two-mass
thresholds has been evaluated by DEM (see Fig.1).
The results are given on pages 2 and 3 and of some of them have been
known before (see 
%discussions in 
\cite{FKV1}). The check of the results 
has been
done by Veretin programs (see discussions in 
\cite{FKV1,FKV2} and references therein).\\
%%%  FIGURE  ===  %%%%%%%%%%%%%%%%
   \begin{figure}[tb]
%\label{fig-2}
\unitlength=1mm
%\begin{picture}(150,100)
%  \put(0,100){%
%   \epsfig{file=tolya.ps,width=100mm,height=150mm,angle=-90}%
\begin{picture}(140,80)
  \put(0,80){%
   \epsfig{file=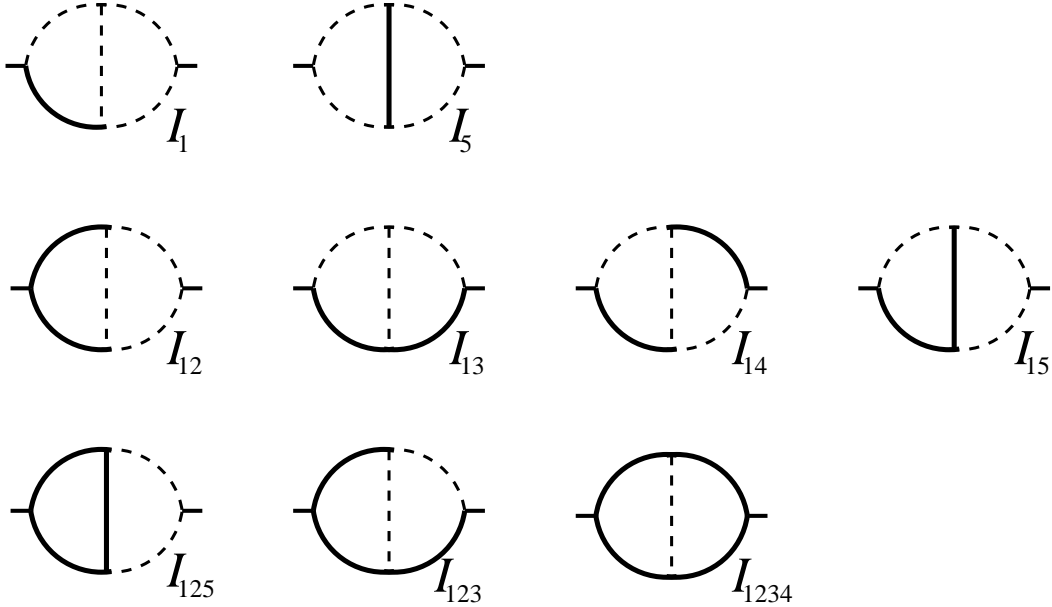,width=80mm,height=140mm,angle=-90}%
}
\end{picture}
%\vskip 0.5cm
 \caption{
%\sl  
Two-loop self-energy diagrams.
Solid lines denote propagators with the mass $m$; dashed lines
denote massless propagators.}
\vskip 0.5cm
 \end{figure}

{\bf b)} The set of three-point two-loop FD with one- and two-mass
thresholds has been evaluated (the results of some of them has been
known before (see \cite{FKV1})) by a combination of DEM and Veretin programs
for calculation of first terms of FD small-moment expansion (see 
discussions in \cite{FKV1,FKV2} and references therein).\\

{\bf 2.} The article \cite{FKK}: \\

The full set of two-point two-loop on-shell master diagrams 
has been evaluated by DEM. The check of the results has been
done by Kalmykov programs (see 
%page 5 and 
discussions in 
\cite{FKK,FKK1} and references therein). \\

{\bf 3.} The articles \cite{GeRede,GeReconf}: \\

The set of three-point and four-point two-loop massless FD 
has been evaluated.\\

\vskip 0.5cm
{\bf Here we demonstrate the results of FD are displayed on Fig.1.}\\

We introduce the notation for
polylogarithmic functions \cite{Lewin}: 
%are particular cases of $S$ functions, namely
\begin{eqnarray}
%\label{Li}
  {\rm Li}_a(z) = S_{a-1,1}(z), ~~~ 
S_{a+1,b}(z) = \frac{(-1)^{a+b}}{a!\,b!}
   \int_0^1 \frac{\log^a(t)\log^b(1-zt)}{t}\,dt.
\nonumber
\end{eqnarray}

  We introduce also the following two variables 
%($z=q^2/m^2$)
\begin{eqnarray}
z=\frac{q^2}{m^2},~~~y=\frac{1-\sqrt{z/(z-4)}}{1+\sqrt{z/(z-4)}}\,.
\nonumber
\end{eqnarray}

  Then\footnote{We would like to note that the
%to note about a similarity of 
coefficients of expansions of the results (\ref{eq}) in respect of $z$ 
%$(Q^2/m^2)^n$ in the below results and and in ones 
are very similar (see \cite{3}-\cite{3.2}) to results 
for the moments of structure functions of deep inelastic scattering,
i.e. to the sums (\ref{1.1}).}

\begin{eqnarray}
q^2 \cdot I_1 \z=\z 
     - \frac{1}{2}\log^2(-z)\log(1-z) 
     - 2\log(-z){\rm Li}_2(z)
     + 3{\rm Li}_3(z) -6 S_{1,2}(z)  
 \nonumber \\   \z-\z 
 \log(1-z) \biggl(  \zeta_2 + 2{\rm Li}_2(z) \biggr)\,, \nonumber \\
\z\z\nonumber\\
q^2 \cdot I_5 \z=\z 
      2 \zeta_2 \log(1+z) + 2\log(-z){\rm Li}_2(-z) + \log^2(-z)\log(1+z)
        + 4\log(1+z){\rm Li}_2(z) 
            \nonumber \\
      \z-\z 2{\rm Li}_3(-z) 
           - 2{\rm Li}_3(z) 
           + 2 S_{1,2}(z^2) - 4S_{1,2}(z) - 4 S_{1,2}(-z)\,, \nonumber \\
\z\z\nonumber\\
q^2 \cdot I_{12} \z=\z
     {\rm Li}_3(z) - 6 \zeta_3 - \zeta_2\log y
     - \frac16\log^3 y - 4\log y\,{\rm Li}_2(y)  
\nonumber\\     \z+\z 
 4{\rm Li}_3(y) - 3{\rm Li}_3(-y) + \frac13 {\rm Li}_3(-y^3)\,,\nonumber \\
\z\z\nonumber\\
q^2 \cdot I_{13} \z=\z 
      - 6 S_{1,2}(z)
      - 2 \log(1-z) \biggl(  \zeta_2 + {\rm Li}_2(z) \biggr)\,, \nonumber \\
\z\z\nonumber\\
q^2 \cdot I_{14} \z=\z  
     \log(2-z) \biggl( \log^2(1-z) -2 \log(-z)\log(1-z) -2{\rm Li}_2(z)
       \biggr)
\nonumber \\  \z-\z 
\frac{2}{3} \log^3(1-z) - 2 \zeta_2 \log(1-z)
\nonumber \\  \z+\z \log(-z)\log^2(1-z)
      \nonumber\\  \z-\z  
S_{1,2}\bigr(1/(1-z)^2\bigl)
     + 2 S_{1,2}\bigr(1/(1-z)\bigl) + 2 S_{1,2}\bigr(-1/(1-z)\bigl)  
+ \frac13 \log^3 y
\nonumber\\
  \z+\z 
%\frac13 \log^3 y    +
\log^2 y\,\biggl( 2 \log(1+y^2) -3 \log(1-y +y^2) \biggr) 
                   \nonumber \\\z-\z 
6 \zeta_3 - {\rm Li}_3(-y^2)
     +\frac23 {\rm Li}_3(-y^3) - 6 {\rm Li}_3(-y) 
         \nonumber \\ 
\z+\z 2\log y\, \biggl( {\rm Li}_2(-y^2) -{\rm Li}_2(-y^3) 
              +3 {\rm Li}_2(-y)    \biggr)\,,  \nonumber \\
\z\z\nonumber\\
q^2 \cdot I_{15} \z=\z 
   2{\rm Li}_3(z) - \log(-z)\,{\rm Li}_2(z) 
              + \zeta_2 \log(1-z) 
\nonumber \\  \z+\z 
    \frac12\log^2 y\,\biggl( 8 \log(1-y) -3 \log(1-y +y^2) \biggr) 
-6 \zeta_3                    \nonumber \\
\z+\z  \frac16 \log^3 y 
- \frac13 {\rm Li}_3(-y^3) + 3 {\rm Li}_3(-y) 
            + 8 {\rm Li}_3(y) 
\nonumber \\ \z+\z 
\log y\, \biggl( {\rm Li}_2(-y^3) -3 {\rm Li}_2(-y) 
              -8 {\rm Li}_2(y)   \biggr)\,, \nonumber \\
\z\z\nonumber\\
q^2 \cdot I_{123} \z=\z 
    - \zeta_2 \biggl(  \log(1-z) +  \log y \biggr)
    - 6 \zeta_3  -\frac32 \log(1-y +y^2)\, \log^2 y 
+ {\rm Li}_3(-y^3) - 9 {\rm Li}_3(-y)
\nonumber \\
    \z-\z  2 \log y\, \biggl( {\rm Li}_2(-y^3) - 3{\rm Li}_2(-y) \biggr)  
      \,, \nonumber \\
\z\z\nonumber\\
q^2 \cdot I_{125} \z=\z  
      -2 \log^2 y\, \log(1-y)
      - 6 \zeta_3 + 6 {\rm Li}_3(y) - 6\log y\, {\rm Li}_2(y)\,, \nonumber \\
\z\z\nonumber\\
q^2 \cdot I_{1234} \z=\z 
    - 6 \zeta_3  - 12 {\rm Li}_3(y) - 24 {\rm Li}_3(-y)
                      \nonumber \\    \z+\z 
8 \log y\, \biggl( {\rm Li}_2(y) + 2{\rm Li}_2(-y) \biggr)  
      + 2 \log^2 y\, \biggl( \log(1-y) 
             \nonumber \\    \z+\z 
 2 \log(1+y) \biggr)\,. 
\label{eq}
\end{eqnarray}

\vskip 1cm

%\vskip 0.5cm
{\bf Here we demonstrate the results of FD are displayed on Fig.2.}\\

\begin{figure}[tb]
\vskip -0.5cm
%\centerline{\vbox{\epsfysize=170mm \epsfbox{joint.eps}}}
\epsfig{figure=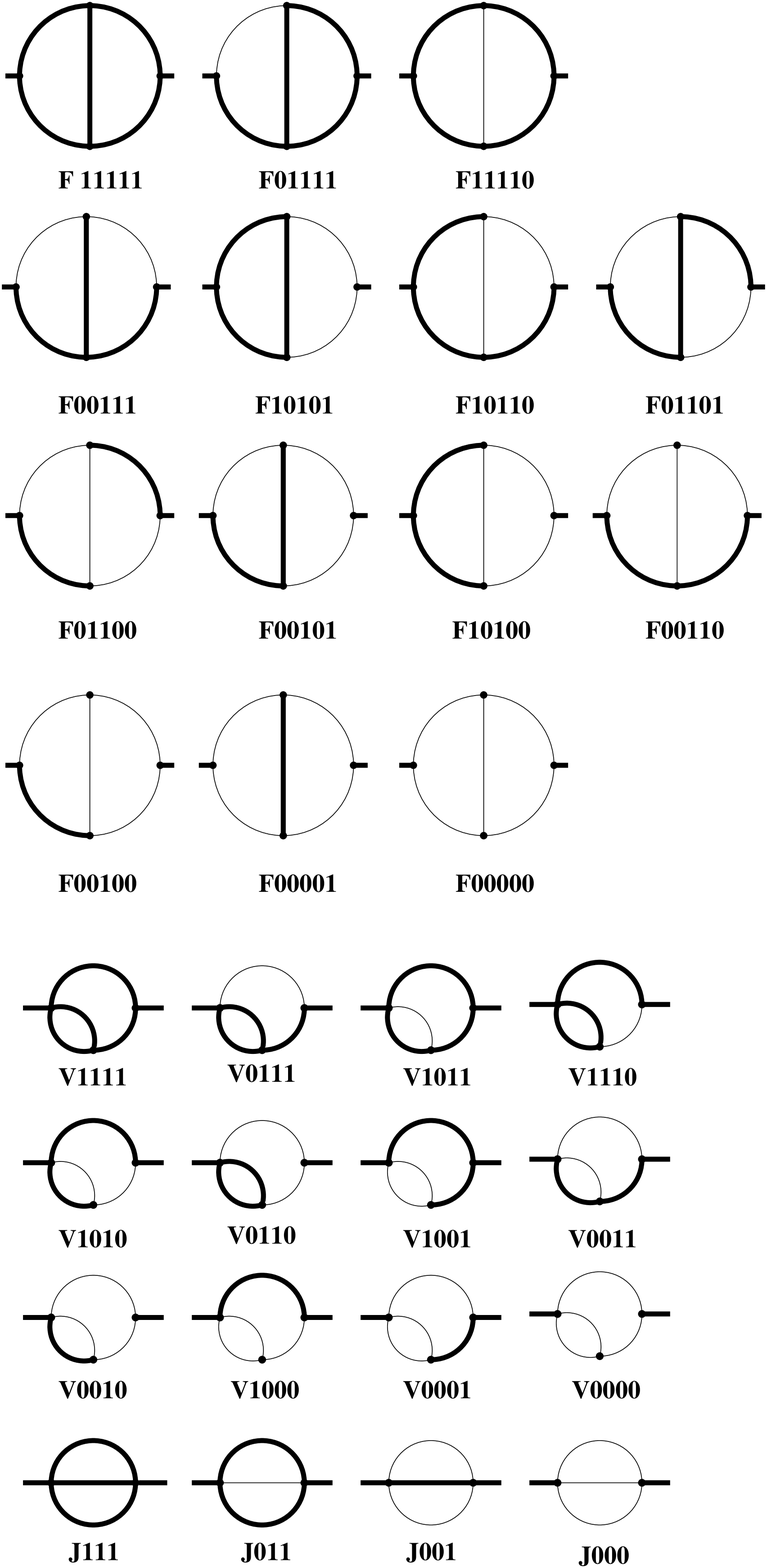,height=7.3in,width=5.2in}
%\begin{picture}(0,170)
%  \put(0,170){%
% \epsfig{file=joint.eps,width=170mm,height=150mm,angle=0}%
%}
%\end{picture}
\caption{\label{set} The full set of two-loop self-energies diagrams
with one mass.  Bold and thin lines correspond to the mass and
massless propagators, respectively.}
\end{figure}

We consider here the following master-integrals in Euclidean 
space-time with dimension $D= 4-2\varepsilon$:

\begin{eqnarray}
 {\bf ONS} \{ {\cal IJ} \} (i,j,m)   & \equiv & K^{-1} 
\int d^Dk P^{(i)}(k,{\cal I}m) 
\left. P^{(j)}(k-p,{\cal J}m) \right|_{p^2=-m^2},
\nonumber \\
{\bf J} \{ {\cal IJK} \} (i,j,k,m)  & \equiv & K^{-2}
\int  d^Dk_1 d^Dk_2
P^{(i)}(k_1,{\cal I} m) P^{(j)}(k_1-k_2,{\cal J}m)
\nonumber  \\
&& ~~~~~
\times 
\left. P^{(k)}(k_2-p,{\cal K}m) \right|_{p^2=-m^2},
\nonumber  \\
{\bf V} \{ {\cal IJKL} \}  (i,j,k,l,m)  & \equiv & K^{-2}
\int  d^Dk_1 d^Dk_2 P^{(i)}(k_2-p,{\cal I}m) 
\nonumber \\
&& 
\times 
P^{(j)}(k_1-k_2,{\cal J}m) P^{(k)}(k_1,{\cal K}m)
\left. P^{(l)}(k_2,{\cal L}m) \right|_{p^2=-m^2},
\nonumber  \\
{\bf F}\{ {\cal ABIJK} \}  (a,b,i,j,k,m)  
& \equiv & m^2 K^{-2}
\int  d^Dk_1 d^Dk_2
P^{(a)}(k_1,{\cal A} m) P^{(b)}(k_2,{\cal B} m)
\nonumber \\
&&
\times
%P^{(b)}(k_2,{\cal B} m)
P^{(i)}(k_1-p,{\cal I}m)
\nonumber \\
&&
\times
P^{(j)}(k_2-p,{\cal J}m)
\left. P^{(k)}(k_1-k_2,{\cal K}m) \right|_{p^2=-m^2},
\nonumber 
\end{eqnarray}

\noindent
where 

$$
K = \frac{\Gamma(1+\varepsilon)}{ \left(4 \pi \right)^{\frac{D}{2}}
\left( m^2 \right)^{\varepsilon}}, ~~~ 
P^{(l)}(k,m) \equiv \frac{1}{(k^2+m^2)^l},$$ 
the normalization factor $1/(2 \pi)^D$ for each loop is assumed,
and ${\cal A,B,I,J,K} = 0,1.$

The finite part of most of the F-type master-integrals can be obtained 
from results of Ref.\cite{FKV1} in the limit $z \rightarrow 1$. 
{\bf F10101} and 
{\bf F11111} have been calculated in Refs.\cite{2david,f11111}, respectively.
Instead of the usually taken {\bf F01101} integral \cite{2david,1david} 
we consider  {\bf J111} as master integral. 
We recall the results of all master integrals for completeness.
The last 
%unknown 
master integral 
%then remains to be 
{\bf F00111} has been found in \cite{FKK}.

The finite part of the integrals of V- and J-type can be found in 
Refs.\cite{short}.
%-\cite{fv3}. 
The calculation of some $\sim \varepsilon$ and
$\sim \varepsilon^2$ parts of master integrals of this type have been 
performed by DEM.
%the differential equation method  \cite{DEM1}-\cite{DEM3}. 

The results for F-type master-integrals are follows:

\begin{equation}
{\bf F}\{ {\cal ABIJK} \} (1,1,1,1,1,m) = a_1 \zeta(3) 
+ a_2 \frac{\pi}{\sqrt{3}} S_2 + a_3 i \pi \zeta(2) + {\cal O} (\varepsilon),
\label{first}
\end{equation}

\noindent
and the coefficients $\{a_i \}$ are given in Table I:

$$
\begin{array}{|c|c|c|c|c|c|c|c|c|} \hline
\multicolumn{9}{|c|}{TABLE ~~~I}    \\   \hline
& {\bf F11111} & {\bf F00111} & {\bf F10101} & {\bf F10110} &
  {\bf F01100} & {\bf F00101} & {\bf F10100} & {\bf F00001}
\\[0.3cm] \hline
a_1 & 1 & 0 & -4 & -1 & 0 & -3 & -2 & -3 \\[0.3cm] \hline
a_2 & -\frac{9}{2} & 9 & \frac{27}{2} & 9 & \frac{27}{2} &
\frac{27}{2} & 9 & 0 \\[0.3cm] \hline
a_3 & 0 & 0 & \frac{1}{3} & 0 & 1 & 1 & \frac{2}{3} & 1
\\[0.3cm] \hline
\end{array}
$$

\noindent
where \cite{S-M,2david,Lewin} 

$$S_2 = \frac{4}{9\sqrt{3}} {\rm Cl}_2 \left(\frac{\pi}{3} 
\right)=0.260434137632 \cdots.$$ 
Here we used the $m^2-i \varepsilon$ prescription. The results for
the remaining master integrals are the following ones:

\begin{eqnarray}
&& 
{\bf V}\{ {\cal IJKL} \} (1,1,1,1,m) = 
\frac{1}{2 \varepsilon^2}
+ \frac{1}{\varepsilon}
\left( \frac{5}{2} - \frac{\pi}{\sqrt{3}} \right)
+ \frac{19}{2} + \frac{b_1}{2} \zeta(2) -4 \frac{\pi}{\sqrt{3}}
- \frac{63}{4} S_2 
%+ \frac{\pi}{\sqrt{3}} \ln 3 
\nonumber \\ 
%&& 
&+& \hskip -8pt
%\hspace*{-8pt} 
\frac{\pi}{\sqrt{3}} \ln 3 +
 \varepsilon \Biggl\{ 
\frac{65}{2} + b_2 \zeta(2) - b_3 \zeta(3)
- 12 \frac{\pi}{\sqrt{3}} - 63 S_2 + b_4 \zeta(2) \ln3
+ \frac{9}{4} b_4 S_2 \frac{\pi}{\sqrt{3}}
\nonumber \\ 
%&& 
&+& \hspace*{-8pt}
\frac{63}{4} S_2 \ln3
+ 4 \frac{\pi}{\sqrt{3}} \ln 3
- \frac{1}{2} \frac{\pi}{\sqrt{3}} \ln^2 3
- \frac{b_5}{2} \frac{\pi}{\sqrt{3}} \zeta (2)
- \frac{21}{2} \frac{{\rm Ls}_3 \left(\frac{2\pi}{3} \right)}{\sqrt{3}}
\Biggr\}
+ {\cal O} (\varepsilon^2),
\nonumber
\end{eqnarray}

\noindent
where the coefficients $\{b_i \}$ are listen in Table II
\footnote{The results for the master integral {\bf V1001} had a little
error (see \cite{DaKa})}:

$$
\begin{array}{|c|c|c|c|c|c|} \hline
\multicolumn{6}{|c|}{TABLE ~~~II}    \\   \hline
            &  b_1 & b_2  & b_3 & b_4 & b_5  \\[0.3cm] \hline
{\bf V1111} &  - 1 & -6 & \frac{9}{2}   & 4 & 9   \\[0.3cm] \hline
{\bf V1001} &    3 &  8 & -\frac{3}{2} & 0 & 21  \\[0.3cm] \hline
\end{array}
$$

\begin{eqnarray}
%&& 
{\bf J111}(1,1,1,m) &=& - m^2 \Biggl(
\frac{3}{2\varepsilon^2} + \frac{17}{4 \varepsilon} + \frac{59}{8}
+ \varepsilon \Biggl\{\frac{65}{16} + 8 \zeta(2) \Biggr \}
\nonumber \\
%&&
&-& 
-\varepsilon^2 \Biggl\{
\frac{1117}{32} - 52 \zeta(2) + 48 \zeta (2) \ln 2 - 28 \zeta (3)
\Biggr\} + {\cal O} (\varepsilon^3)
\Biggr), 
%\nonumber \\
\label{master-j111} \\
& & \nonumber \\
{\bf J011}(1,1,2,m) &=& \frac{1-4\varepsilon}{2 (1-2\varepsilon) 
(1-3\varepsilon)}
\Biggl( \frac{1}{\varepsilon^2} + 2 \frac{\pi}{\sqrt{3}}
- \frac{2}{3} \zeta(2) 
\nonumber \\
%&& 
&+& \varepsilon \Biggl\{
8 \frac{\pi}{\sqrt{3}} - \frac{2}{3} \zeta(2)
-6  \frac{\pi}{\sqrt{3}} \ln 3
+ \frac{2}{3} \zeta(3)
+ 27 S_2 \Biggr\}  
+ {\cal O} (\varepsilon^2)
\Biggr),
\label{master-j011-1}\\
%\end{eqnarray}
& & \nonumber \\
{\bf J011}(1,1,1,m) &=& -  \frac{m^2}{2} 
\frac{4- 15 \varepsilon}{(1-2\varepsilon) (1-3\varepsilon) (2-3\varepsilon)}
\Biggl( \frac{1}{\varepsilon^2} + \frac{3}{2} \frac{\pi}{\sqrt{3}}
\nonumber \\
&+& \varepsilon \Biggl\{
\frac{45}{8} \frac{\pi}{\sqrt{3}} 
-  \frac{9}{2}  \frac{\pi}{\sqrt{3}} \ln 3
+ \frac{81}{4} S_2 \Biggr\}  \nonumber \\
%&& 
&+&
 \varepsilon^2 \Biggl\{
12 - \zeta(2) - \frac{1863}{16} S_2 -\frac{867}{32} \frac{\pi}{\sqrt{3}} 
+ \frac{207}{8} \frac{\pi}{\sqrt{3}} \ln 3 
+ \frac{243}{4}S_2 \ln 3
\nonumber \\
%&& 
&-& 
\frac{27}{4} \frac{\pi}{\sqrt{3}} \ln^2 3
- 21 \frac{\pi}{\sqrt{3}} \zeta(2)
- \frac{81}{2}  \frac{{\rm Ls}_3 \left(\frac{2\pi}{3} \right)}{\sqrt{3}} 
\Biggr\} +
{\cal O} (\varepsilon^3)
\Biggr),
\label{master-j011-2}\\
%\end{eqnarray}
& & \nonumber \\
%\begin{eqnarray}
{\bf ONS11}(1,1,m) &=&
\frac{1}{1-2\varepsilon} \Biggl[
\frac{1}{\varepsilon}  - \frac{\pi}{\sqrt{3}}
+ \varepsilon \left\{ \frac{\pi}{\sqrt{3}}\ln3 - 9 S_2 \right\}
%\nonumber 
\label{master-ons11}\\
&+& \varepsilon^2 \left\{
9 S_2 \ln3 - \frac{1}{2}\frac{\pi}{\sqrt{3}} \ln^2 3
- 6 \frac{{\rm Ls}_3 \left(\frac{2\pi}{3} \right)}{\sqrt{3}}
- 3 \frac{\pi}{\sqrt{3}} \zeta (2) \right\}
+ {\cal O} (\varepsilon^3) \Biggr],
\nonumber
%\label{master-ons11}
\end{eqnarray}

\noindent
where \cite{Lewin}  
$$ {\rm Ls}_3(x) = -\int_0^x \ln^2 
\left| 2 \sin \frac{\theta}{2}\right| d \theta ~~\mbox{ and }~~
{\rm Ls}_3 \left(\frac{2\pi}{3} \right) = -2.14476721256949 \cdots$$

   The above results were checked numerically. Pad\'e approximants 
were calculated from the small momentum Taylor expansion of the
diagrams \cite{small}. The Taylor coefficients were obtained by means of the
package \cite{TLAMM} with the master integrals taken from \cite{small1}.
Further we made use 
of the idea of Broadhurst \cite{numeric} to apply the FORTRAN program 
{\bf PSLQ}
\cite{pslq} to express the obtained numerical values in terms of
a `basis' of irrational numbers, which were predicted by  DEM.

Let us point out that the numbers we obtain are related to
polylogarithms at the sixth root of unity\footnote{For 
the results obtained in $1/N$ expansion, however, the 
arguments of polylogarithms 
%seem to 
have other values (see \cite{BGK,BrKo,KoLi,KoLi1}).}
and hence are in the same class of
transcendentals obtained by Broadhurst \cite{numeric}
in his investigation of three-loop diagrams
which correspond to a closure of the two-loop topologies considered here.\\

  {\large {\bf Acknowledgments.}}\\

Author would like to express his sincerely thanks to the Organizing
Committee of the PNPI Winter School for the kind invitation, 
%and 
the financial support,
%at  such remarkable Conferences, 
and for fruitful discussions.

Author is supported in part by INTAS grant 00-366.
He thanks also the Alexander von Humboldt Foundation for its support
at the beginning of the study.

\end{document}